\documentclass{elsart}
\usepackage{lineno}

\usepackage{graphicx}           
\usepackage{bm}                 
\usepackage{amsmath}            
\usepackage{amsfonts}
\usepackage{amssymb}
\usepackage{latexsym}           

\usepackage[square,comma,numbers,sort&compress]{natbib}
\newcommand{\ket}[1]{\mbox{$|#1\rangle$}}

\begin{document}
\begin{frontmatter}

\title{Optimal computation with non-unitary quantum walks}

\author[label1]{Viv Kendon},
\ead{V.Kendon@leeds.ac.uk}
\author[label1,label2]{Olivier Maloyer}

\address[label1]{School of Physics and Astronomy, University of 
Leeds, LS2 9JT, United Kingdom.}
\address[label2]{Magist{\`e}re de Physique Fondamentale d' Orsay, Universit\'{e} Paris-Sud, Orsay, France.}

\begin{abstract}
Quantum versions of random walks on the line and the cycle show a quadratic
improvement over classical random walks in their spreading rates and
mixing times respectively.  Non-unitary quantum walks can provide a
useful optimisation of these properties, producing a more uniform
distribution on the line, and faster mixing times on the cycle.
We investigate the interplay between quantum and random dynamics
by comparing the resources required, and examining numerically
how the level of quantum correlations varies during the walk.
We show numerically that the optimal non-unitary quantum walk
proceeds such that the quantum correlations are nearly all removed
at the point of the final measurement.  This requires only
$O(\log T)$ random bits for a quantum walk of $T$ steps.
\end{abstract}


\end{frontmatter}


\section{Introduction}
\label{sec:intro}

Quantum computing has produced a range of algorithms showing 
improvements over classical algorithms, among the most celebrated
are Grover's search of an unsorted database \cite{grover97a} and Shor's
algorithm for factoring large numbers \cite{shor95a}.
In the search for new quantum algorithms, quantum versions of Markov
chains are a natural place to look, since
classical Markov chains provide the basis for some of the best known
algorithms for hard problems such as approximating the permanent of
a matrix \cite{jerrum01a} and $k$SAT (with $k>2$) \cite{schoning99a}.

Simple quantum generalisations of classical random walks 
spread quadratically faster on the line \cite{ambainis01a,nayak00a},
and mix quadratically faster on the cycle \cite{aharonov00a}.
These promising early results were soon followed by 
several algorithms based on quantum walks. 
Shenvi et al \cite{shenvi02a} proved a quantum walk
can solve the unsorted database search problem quadratically faster,
and Childs et al \cite{childs02a} proved an exponential 
speed up for crossing a particular type of graph.  Ambainis
\cite{ambainis03a} gives an overview of quantum walk algorithms, and
Kempe \cite{kempe03a} provides an introductory review
of quantum walks and their properties.

Quantum walks on simple one-dimensional structures remain a fertile
testing ground for further research.
It was shown numerically by Kendon and Tregenna \cite{kendon02c}
and recently proved by Richter \cite{richter06a,richter06b} that the
addition of random noise or measurements to the quantum walk dynamics can
optimise the spreading and mixing properties for quantum walks on
both the line and the cycle.  A detailed survey of the effects of 
such decoherence in quantum walks can be found in \cite{kendon06b}.
In this paper we examine how the interplay between quantum evolution and
random noise or measurements produces optimal computational properties.
The paper is organised as follows.  In \S\ref{sec:qrwdef} we describe
the basic quantum walk model we used for this study, and in \S\ref{sec:qrwnonu}
we extend this model to include non-unitary operations.
We then outline how to count both quantum and random computational
resources in a comparable way in \S\ref{sec:counting}.
Section \ref{sec:negativity} introduces the measure of quantum
correlations we are using, the negativity. In \S\ref{sec:linent}
we present our results for the quantum walk on the line, and in
\S\ref{sec:cyclent} our results for the walk on the cycle.  Discussion
and conclusions are given in \S\ref{sec:conc}.

\section{Quantum walks on the line and the cycle}
\label{sec:qrwdef}

This study considers quantum walks taking place in discrete time and space,
using a quantum coin to control the choice of direction.  This is
because we measure how ``quantum'' the walk is by examining the
correlations between the quantum coin and the quantum walker.  
A different method for assessing ``quantumness'' would be needed for
continuous-time quantum walks (introduced for quantum algorithms by
Farhi and Gutmann \cite{farhi98a}).

A discrete-time (coined) quantum walk dynamics consists of a quantum
``coin toss'' operation $\mathbf{C}$, followed by a shift operation $\mathbf{S}$ to move the 
quantum walker to a new position.  These are repeated alternately
for $T$ steps of the quantum walk, and the final position of the
quantum walker is measured.
For quantum walks on the line and the cycle, we have just two
choices of which way to step, so the quantum coin is two-sided.  
We write $|x,c\rangle$ for a quantum walker at position $x$ with a coin
in state $c \in \{+1,-1\}$.  For a walk on the line, $x \in \mathbb{Z}$
and for a walk on a cycle of size $N$, we have $x\in\mathbb{Z}_N$.
A classical random walk can only occupy one location at any given time, but
a quantum walker can be in a superposition of different locations.
The full state of the quantum
walk $|\Psi(t)\rangle$ can be written as a combination of terms in each
basis state $|x,c\rangle$,
\begin{equation}
|\Psi(t)\rangle = \sum_{x,c}\psi_{x,c}(t)|x,c\rangle,
\label{eq:state}
\end{equation}
where $\psi_{x,c}(t) \in \mathbb{C}$. By convention, the normalisation 
is defined to be
\begin{equation}
\sum_{x,c}|\psi_{x,c}(t)|^2 = 1.
\end{equation}
When the quantum walk is measured (in the basis just defined),
the walker is found in a single location with a definite coin
state, but we cannot predict with certainty which state this
will be, it could be any of the states in the superposition described by 
eq.~(\ref{eq:state}).  The probability of finding the quantum walker at
position $y$ with the coin in state $b$ is given by
\begin{equation}
P(y,b,t) = |\langle y,b|\Psi(t)\rangle|^2 = |\psi_{y,b}(t)|^2
\end{equation}
since the basis states are orthogonal, $\langle y,b|x,c\rangle = \delta_{xy}
\delta_{bc}$.

The coin toss and shift operators can be defined in terms of their
action on the basis states $|x,c\rangle$,
\begin{equation}
\mathbf{S}|x,c\rangle = |x+c,c\rangle
\end{equation}
\begin{equation}
\mathbf{C}|x,c\rangle = (|x,-c\rangle + c|x,c\rangle)/\sqrt{2}
\end{equation}
One can add more general bias or phase into the coin toss operation, see,
for example, \cite{bach02a}, but this does not greatly change the basic
properties of the quantum walk on a line or cycle, so we will consider
only the unbiased case in this paper.
For a quantum walk starting at position $x=0$ with the coin in a
superposition state $(|{-1}\rangle + i|{+1}\rangle)/\sqrt{2}$,
(where $i=\sqrt{-1}$)
we can write a quantum walk of $T$ steps as
\begin{equation}
|\Psi(T)\rangle = (\mathbf{SC})^T(|0,-1\rangle + i|0,+1\rangle)/\sqrt{2}
\end{equation}
The solution for $|\Psi(T)\rangle$ may be obtained by various methods
such as Fourier analysis \cite{nayak00a} and path counting \cite{ambainis01a},
and has been studied extensively.  The key result 
is that spreading on the line proceeds linearly with the
number of time steps.
We can use the standard deviation $\sigma_Q(T)$ of the probability distribution
to quantify the spreading rate.  For a quantum walk on the line,
\begin{equation}
\sigma_Q(T) = \sum_{x,c}x^2P(x,c,T)\simeq\left(1-\frac{1}{\sqrt{2}}\right)^{1/2}T,
\end{equation}
asymptotically in the limit of large $T$.  In contrast, for the
classical random walk, the standard deviation is $\sigma_C(T) = \sqrt{T}$.

On the cycle, we are interested in mixing times rather than
spreading.  Mixing times can be defined in a number of different
ways, we choose the definition given in \cite{aharonov00a},
\begin{equation}
M(\epsilon)=\min\left\{T\enspace|\enspace\forall\enspace t>T: ||P(x,t)-P_u||_{\text{tv}}<\epsilon\right\}
\label{eq:mixdef}
\end{equation}
where $P_u$ is the limiting  distribution over the cycle, and
the total variational distance (TVD) is defined as
\begin{equation}
||P(x,T) - P_u||_{\text{tv}} \equiv \sum_x|P(x,T) - P_u|.
\label{eq:tvd}
\end{equation}
A classical random walk on the cycle mixes to within
$\epsilon$ of the uniform distribution in time
proportional to $N^2\log(1/\epsilon)$, 
where $\epsilon$ can be chosen arbitrarily small.

Pure quantum walks, on the other hand, do not mix to a stationary distribution.
Their deterministic dynamics ensures they continue to oscillate indefinitely.
There are several ways to obtain mixing behaviour,
first explored by Aharonov et al \cite{aharonov00a}.
By defining a time-averaged probability distribution for the quantum walk,
\begin{equation}
\overline{P(x,c,T)} = \frac{1}{T}\sum_{t=0}^{T-1}P(x,c,t)
\label{eq:meanP}
\end{equation}
they proved that $\overline{P(x,c,T)}$ does converge to a stationary
distribution on a cycle, and that on odd-sized cycles the stationary
distribution is uniform.
A mixing time can be defined for $\overline{P(x,c,T)}$,
\begin{equation}
\overline{M}(\epsilon)=\min\left\{T\enspace|\enspace\forall\enspace t>T: ||\overline{P(x,t)}-P_u||_{\text{tv}}<\epsilon\right\}.
\label{eq:meanmixdef}
\end{equation}
and Aharonov et al \cite{aharonov00a} proved that, for odd-sized cycles,
$\overline{M}(\epsilon)$ is bounded above by $O(\epsilon^{-3}N\log N)$,
almost quadratically faster (in $N$) than a classical random walk.
Kendon and Tregenna \cite{kendon02c},
observed numerically that $\overline{M}(\epsilon)\sim O(N/\epsilon)$, 
this has been recently confirmed analytically by Richter \cite{richter06b}.

Notice that we pay a price for our time-averaging: the scaling with
the precision $\epsilon$ is now linear instead of logarithmic.
Aharonov et al \cite{aharonov00a} provide a fix for this in the form of a
``warm start''.  The quantum walk is run several times, each repetition
starting from the final state of the previous run.  A small number of such
repetitions is sufficient to reduce the scaling of the mixing time
$\overline{M}(\epsilon)$ to logarithmic in $\epsilon$.

Both the quantum walk on the line and the cycle thus provide
a quadratic speed up over classical random walks.
This quadratic speed up does not carry over to all quantum walks on
higher dimensional structures, see, for example,
\cite{mackay01a,tregenna03a,krovi06a}.  It remains an open question how
ubiquitous this behaviour really is.

\section{Non-unitary quantum walks}
\label{sec:qrwnonu}

The quantum walk dynamics described in the previous section
is a pure quantum evolution terminated by a final
measurement to determine the outcome of the quantum walk.
If, instead, the quantum walk is measured after every step,
it is easy to see that it becomes a classical random walk.
In between these two extremes, a smaller number of
measurements can be included in the quantum walk dynamics,
as in the ``warm start'' already mentioned for the walk on the cycle.
This is not a new idea.  Quantum walks with measurements were
first explored in \cite{aharonov92a}
for controlling a physical quantum walk in an atom optical system.  

We are thus led to define a more general quantum walk by including
the possibility of non-unitary operations at each step
(pure quantum dynamics is unitary, measurements are non-unitary).
Kendon and Sanders \cite{kendon04a} provide a general formulation
of non-unitary quantum walks using superoperators.  We will not need the
full generality here, we restrict this study to randomly-occurring uncorrelated
non-unitary events, and measurements at specific chosen times during the
quantum walk.  Nonetheless, the evolution of the quantum walk
must now be described using a density operator $\bm\rho(t)$ given by
\begin{equation}
\bm\rho(t+1) = (1-p)\mathbf{SC}\bm\rho(t)\mathbf{C}^{\dag}\mathbf{S}^{\dag}
          + p\sum_j\mathbb{P}_j\mathbf{SC}\bm\rho(t)
                \mathbf{C}^{\dag}\mathbf{S}^{\dag}\mathbb{P}_j^{\dag}.
\label{eq:decdyn}
\end{equation}
Here $\mathbb{P}_j$ is a projection that represents the action of the
non-unitary operator and $p$ is the probability of applying this operator
per time step, or, completely equivalent mathematically,
to a weak coupling between the quantum walk system and a
Markovian environment with coupling strength $p$.
For a pure state, the density operator $\bm\rho(t)\equiv |\Psi(t)\rangle
\langle\Psi(t)|$, it thus has the normalisation $\rm{Tr}[\bm\rho(t)]=1$.

Full analytical solution of a non-unitary quantum walk has been
done only for a few instances.  For quantum walks on the line, Brun et al
\cite{brun02c} analysed the case of random measurements on the coin only.
While analytical solution is challenging, eq.~(\ref{eq:decdyn})
lends itself readily to numerical simulation since
$\bm\rho$, $\mathbf{S}$ and $\mathbf{C}$ can be manipulated as complex
matrices, while the $\mathbb{P}_j$ generally remove some or all
of the off-diagonal entries in $\bm\rho$.
Kendon and Tregenna \cite{kendon02c} evolved eq.~(\ref{eq:decdyn})
numerically for various choices of $\mathbb{P}_j$,
projection onto the position space, projection into the coin space
in the preferred basis ($\ket{\pm 1}$), and projection of both
coin and position.  In all cases, the spreading
rate is reduced, in the long time limit \cite{brun02c},
it becomes proportional to $\sqrt{T}$ instead of proportional to $T$.
More interesting behaviour is seen for intermediate times
and noise rates $p$ with noise applied to the position.
Kendon and Tregenna observed that, for $2\lesssim pT \lesssim 5$,
the distribution becomes very close to uniform while retaining the
full quantum linear spreading rate \cite{kendon02c}.
With noise applied to the coin only, the distribution retains a
cusp shape.
To quantify this the TVD given by eq.~(\ref{eq:tvd}) is used,
this time with $P_u$ defined to be a top-hat of appropriate width
$x\in\{\pm T/\sqrt{2}\}$, see \cite{nayak00a,ambainis01a}.

For a quantum walk on the cycle subjected to Markovian noise, the mixing 
behaviour is dramatically improved, provided noise is applied to the position.
The noise guarantees mixing to the
uniform distribution, and a similar judicious choice of
$2\lesssim pN \lesssim 5$ produces the minimum mixing time \cite{kendon02c}.
Noise on the coin only does cause the quantum walk on a cycle to mix,
but not significantly faster than a classical random walk.
Furthermore, the time-averaging is no longer needed, fast mixing
occurs in time $O(N\log(1/\epsilon)$, as shown in numerical work by
Maloyer and Kendon \cite{kendon06b,maloyer06a}, and
recently proved by Richter \cite{richter06b}.
Significantly, Richter also proved that the randomness produced by quantum
measurements alone is sufficient to produce an optimal mixing time.
By applying measurements at regular intervals, instead of randomly with
probability $p$, the speed up is still obtained.

We thus have two examples where the optimal computational properties are
obtained for a judicious combination of quantum dynamics and measurements,
which introduce a component of randomness into the otherwise deterministic
quantum dynamics.  In order to understand more about how this mixture of quantum
and random resources combine to produce their computational power, we next
describe how we can compare them in the quantum walk.

\section{Comparing quantum and classical computational resources}
\label{sec:counting}

Our goal is to compare the resources required to perform a quantum
walk with those required for a classical random walk, to find out
which is more efficient when the random resources are 
taken into account as well as the memory and gate operations.
What we will find is that the optimal non-unitary quantum walk 
on the line run for $T$ steps requires only $O(\log T)$ extra
quantum gates and ancillae compared with a pure quantum walk.
Since the pure quantum walk uses $O(T\log T)$ quantum gates and
$O(\log T)$ qubits, the extra resource costs are insignificant
to leading order.  A classical random walk producing the same
computational outcome requires $T^2$ steps.  This can be accomplished
using $O(T^2\log T)$ quantum gates and $O(T^2)$ ancillae.
The extra resources compared with a fully quantum walk
are mainly needed to generate the random bits.
Applying the same reasoning to the optimal quantum walk on a cycle shows
that it is similarly efficient.  We now explain in detail how
these results are obtained.

We can quantify the randomness added to, or generated by, the
non-unitary quantum walk by counting up the number of random bits, 
either supplied to time the random measurements, or produced from the
results of those measurements and then used as input to the next
iteration of the quantum walk.  The quantum resources can be
quantified by counting the number of gate operations required, in
a similar manner to classical computational complexity.  What we
then need is a way to compare the quantum and random resources.

We first observe that, if we have only quantum resources to hand,
we can generate random bits by using a quantum gate followed by a
measurement.  To illustrate, using a quantum coin by itself,
we apply the coin toss then measure.  For example,
\begin{equation}
\mathbf{C}|{+1\rangle} = \{|{-1}\rangle + |{+1}\rangle\}/\sqrt{2}.
\end{equation}
The measurement outcome is $|{-1}\rangle$ 50\% of the time, and
$|{+1}\rangle$ the other 50\% of the time, i.e. one random bit.
We can repeat this procedure (starting with the coin in whichever
state was obtained after measuring) to obtain further random bits.
Of course, the converse does not work: if we have only classical random
bits, we cannot efficiently simulate all the properties of
a general quantum system.  What is notable is that a quantum computer
comes with randomness as a ``built in'' function, whereas classical
computation does not and it must be added separately, or faked
(pseudo-randomness) at some significant computational cost.

An alternative way of viewing non-unitary quantum evolution is as 
a pure quantum evolution in a larger system that includes an ancilla or
environment degrees of freedom.  Let us illustrate
this with the quantum coin in an arbitrary state
\begin{equation}
|\Psi_c\rangle = \alpha|{-1}\rangle - \beta|{+1}\rangle.
\label{eq:qubit}
\end{equation}
If we measure in the computational basis, we find the coin in
state $|{-1}\rangle$ with probability $\alpha^2$ and in state
$|{+1}\rangle$ with probability $\beta^2$.  Now we add an ancilla qubit,
another two state system with states $|e_0\rangle$ and $|e_1\rangle$.
We start with the ancilla in state $|e_0\rangle$ and apply a
CNOT gate with the quantum coin as the control and the ancilla as the target,
\begin{equation}
\text{CNOT}|\Psi_c\rangle|e_0\rangle 
	= \alpha|{-1}\rangle|e_0\rangle - \beta|{+1}\rangle|e_1\rangle.
\end{equation}
Now we discard the ancilla, and consider the state of the quantum coin
only.  Mathematically, we trace out the ancilla, leaving
\begin{equation}
\text{Tr}_e\left[\alpha|{-1}\rangle|e_0\rangle
	- \beta|{+1}\rangle|e_1\rangle\right]
	= \alpha^2|{-1}\rangle\langle{-1}| + \beta^2|{+1}\rangle\langle{+1}|.
\end{equation}
This is now a mixed state with classical probabilities that match the
probabilities measured in the quantum state eq.~(\ref{eq:qubit}).
However, it now has no useful quantum properties.
Notice in particular that the phase 
($-\beta$ rather than $+\beta$) is lost from the original superposition
state.  The quantum phases are crucial for generating the computational
speed up in a quantum walk, via the cancellations (interference) they
produce when the quantum walk arrives at the same location via two
different paths.  Kendon and Sanders \cite{kendon04a} employ this method
of measuring the quantum coin with a more general coupling to explore
the effects of weak measurements.

Based on these examples, we can see that one way to count resources in a
non-unitary quantum walk is to count them all as quantum resources. 
We count the number of quantum gates, and we also tally up the number
of quantum bits (qubits) we require, including the ancillae that we
use in the measurement process.  A random bit thus requires one quantum gate,
and one qubit ancilla to couple to the qubit being measured.  Note that we
cannot ``recycle'' the ancillae, since returning them to a pure quantum state
$|e_0\rangle$ would require further operations and ancillae.

We next apply the quantum resource counting to the
quantum walk on the line, run for $T$ steps with non-unitary noise rate $p$.
This is equivalent to including a total of $pT$ measurements
at random intervals during our quantum walk.
Since the noise can be applied as a weak coupling at every step, we do
not need to generate random numbers to decide when to apply the measurements.
Each step of the quantum walk uses two quantum gates,
$\mathbf{C}$ and $\mathbf{S}$.
Now, $\mathbf{C}$ acts non-trivially on the coin qubit only, but
$\mathbf{S}$ acts on the whole quantum walk system.  We assume the position
of the quantum walker is encoded in a quantum register of size
$\lceil\log(2T+1)\rceil$
qubits.  Thus $\mathbf{S}$ will need to consist of $O(\log T)$
elementary gates each acting on one or two qubits.
Next we turn to the noise.  There are two cases we need to consider: noise
acting on the coin only, and noise acting on the position space as well.
(Noise acting on both the position and coin differs only $O(1)$ in
resources required from noise acting on the position only, so we will
consider these two cases as one.)
We have already noted that noise on the coin only does not have the same
computational effect as noise acting on the position.
We will now see that the required resources are different.
The $pT$ random measurements applied to the coin only require 
$O(pT)$ quantum gates, and $O(pT)$ ancillae.
Applying noise to the position requires
$O(pT\log T)$ quantum gates and $O(pT\log T)$ ancillae.
These random bits are, of course, drawn from the distribution of the
quantum walk on the line, which is not uniformly random, so this estimate
is an upper bound on the true minimum required\footnote{
The distribution of the quantum walk on the line is fairly close to
uniform \cite{nayak00a}, so this will be a good estimate of the requirements}.
The number of qubits we need for the quantum walk is one for the coin plus
$\lceil\log(2T+1)\rceil$ for the position.
We thus require a total of\\
\begin{tabular*}{\textwidth}{@{\extracolsep{\fill}}c@{}c@{}c@{}}
		& coin noise		& position noise \\
quantum gates:	& $O(T\log T + pT)$	& $O(T\log T + pT\log T)$ \\ 
qubits:		& $O( \log T + pT)$	& $O( \log T + pT\log T )$ \\
\end{tabular*}

Let us check the extreme cases, pure quantum and fully classical.
For the pure quantum walk (no noise or randomness) we have $p=0$,
so for large $T$ we require
$O(T\log T)$ quantum gates and $O(\log T)$ qubits.
For the classical random walk, $p=1$ corresponding to measurements applied
at every step, but it is sufficient to apply measurements to the coin only,
so we find we still require $O(T\log T)$ quantum gates, but we need an
additional $T$ ancillae to generate the randomness at each step.
However, to make a fair comparison in terms of
the outcome of the walks, the classical random walk must run for longer.
To achieve the same spreading as a quantum walk run for $T$
steps, we require $\sim T^2/2$ steps of a classical random walk,
requiring $O(T^2\log T)$ quantum gates and $O(T^2)$ ancillae.
Viewed from this perspective, randomness is an expensive resource!
Of course, what we did not count is the cost of maintaining the quantum
coherences necessary for the pure quantum walk to function properly.
However, this depends on the particular physical implementation of a
quantum computer.  While there is no fundamental minimum requirement
that we know of, in practice we expect the costs of maintaining
coherence in a realistic quantum computer to be very high in terms of
the quantum error correction overhead that will be required
\cite{gottesman97a}.

Finally we are ready to consider the optimal quantum walk.
As already noted, the closest to uniform distribution requires
measurements to be applied to the position, and is obtained
for a noise rate such that $pT\sim O(1)$.
Our resource requirements under this condition are
$O(T\log T + \log T)$ quantum gates and $O((1 + O(1))\log T)$ qubits.
We thus need to add only $O(\log T)$ extra quantum gates and ancillae,
to optimise the quantum walk, a small increase in the resources compared
to the total resources required for the pure quantum walk.

We can do the same comparisons for the non-unitary quantum walk on the cycle.
To obtain a distribution within $\epsilon$ of the uniform distribution on
a cycle of size $N$, we run for $M(\epsilon)$ steps with noise rate
rate $p$.\\
\begin{tabular*}{\textwidth}{@{\extracolsep{\fill}}c@{}c@{}c@{}}
		& coin noise		& position noise \\
quantum gates:	& $O(M(\epsilon)\log N + pM(\epsilon))$	& $O(M(\epsilon)\log N + pM(\epsilon)\log N)$ \\ 
qubits:		& $O( \log N + pM(\epsilon))$	& $O( \log N + pM(\epsilon)\log N )$ \\
\end{tabular*}

For the time-averaged mixing time $\overline{M(\epsilon)}$, we need to add
$O(\log\overline{M(\epsilon)})$ gates and qubits, to choose a random time
at which to stop between $0\dots T$,
as per the definitions in eqs.~(\ref{eq:meanP}) and (\ref{eq:meanmixdef}). 

Using $M(\epsilon)\sim O(N\log(1/\epsilon))$ for the quantum walk with noise
rate $p$ such that $pN\sim O(1)$, and $M(\epsilon)\sim O(N^2\log(1/\epsilon))$
for a classical random walk, we can compare the resources required in each case.
Noise on the coin applied at every step ($p=1$) is sufficient to produce
a classical random walk, which thus requires $O(N^2\log N\log(1/\epsilon))$
quantum gates and $O(N^2\log(1/\epsilon))$ qubits.  For the optimal quantum
walk we consider only noise on the position, since noise on the coin only does
not allow the quantum walk to mix significantly faster than a classical
random walk \cite{kendon02c}.
A pure quantum walk run for $N\log(1/\epsilon)$ steps (which does not mix)
requires $O(N\log N\log(1/\epsilon))$ quantum gates and $O(\log N)$ qubits.
For optimal mixing we need to add $O(\log N\log(1/\epsilon))$ quantum gates and
qubits.
Again, only a small amount of randomness is required to optimise the quantum
walk on a cycle, while the cost for the classical random walk is much higher.

\section{Entanglement in mixed states}
\label{sec:negativity}

Having estimated the number of quantum gates required for the
quantum walk, it would be interesting to know whether they are
actually used to full effect.  The unitary operations that
generate quantum correlations are not guaranteed to do so
every time they are applied. 
Depending on the current state of the system, they can just as easily
remove correlations as add them.  So, we will look directly at the
quantum correlations in the quantum walk system
by calculating the entanglement between the coin and
the position of the quantum walker.
We choose our entanglement measure to be the negativity because this can 
be calculated numerically in a fairly straightforward manner for density
operators such as $\bm\rho(t)$, and there are few options that meet this
criterion.  
This is not ideal.  The negativity does not relate directly to
information-theoretic quantities, like entropy, that
could be compared with the number of random bits,
but it will suffice for our purposes in this study.

The negativity is defined as follows.  First we must choose
a division of our system into two (or more) subsystems between which to
identify the entanglement.  For our quantum walk, the natural division is
between the coin and the quantum walker's position.  We note that the
entanglement across this division will be the same whether we regard the
quantum walk as a qubit coin and a unary position, or as the position
encoded in a binary quantum register (because the Hilbert spaces are the
same size).  We perform a
partial transpose on one subsystem to obtain a new matrix 
$\bm\rho'(t)$.  For example, the partial transpose with respect to
the coin subsystem is 
\begin{equation}
\rho'_{xc,yb}(t) = \rho_{xb,yc}(t)
\end{equation}
where $x$, $y$ are position indices and $c$, $b$ are coin state indices.
Next, we determine the spectrum of $\bm\rho'(t)$, denoted by$\{\lambda'_i\}$.
The normalisation of $\bm\rho(t)$ is carried over to $\bm\rho'(t)$, so
$\sum_i\lambda'_i = 1$, but unlike $\bm\rho(t)$, it is possible for
$\bm\rho'(t)$ to have negative eigenvalues.  The negativity is defined 
\cite{peres96a,horodecky96a,vidal01a} as
\begin{equation}
E=\frac{1}{2}\left(\sum_i|\lambda'_i| -1\right),
\end{equation}
which is just the sum of the negative eigenvalues.
The negativity ranges between zero and one, with any non-zero value
indicating entanglement is present.  If the negativity is zero, it means the
state is probably not entangled, but there can be exceptions
\cite{horodecky98a}.  The exceptions are known to be relatively rare
in the set of all possible states \cite{zyczkowski99a}, and the
entanglement they contain is difficult to apply to useful quantum tasks
\cite{horodecky98a}.  For this study, we will not need the fine-grained
detail of these possible exceptions.

\section{Entanglement in a quantum walk on the line}
\label{sec:linent}

The entanglement in a pure state quantum walk has been studied previously,
see for example, \cite{carneiro05a}.  It fluctuates with each step, and
eventually settles down to an asymptotic value that depends on the initial
state of the quantum coin, and on any bias in the quantum coin operator
$\mathbf{C}$.  The addition of noise smooths out this behaviour, see
\begin{figure}
    \begin{center}
	\resizebox{0.6\columnwidth}{!}{\includegraphics{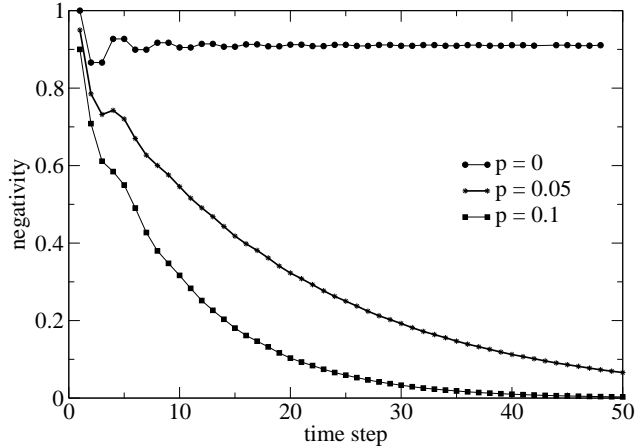}}
    \end{center}
    \caption{Negativity for noise rate $p=0$ (circles), $p=0.05$
	(asterisks) and $p=0.1$ (squares) for a quantum walk on the
	line with noise applied to both coin and position, run for
	50 time steps.}
    \label{fig:ld50neg}
\end{figure}
fig.~\ref{fig:ld50neg}, and steadily reduces the level of entanglement
between the coin and the position.
Recall that a noise rate $p$ can be interpreted as a weak measurement of
strength $p$ applied every step, or, a full measurement applied only
with probability $p$, whereupon the steady reduction in entanglement can
be regarded as the average effect of a few random measurements.
Our simulations simply take eq.~(\ref{eq:decdyn}) and evolve it
numerically, calculating the negativity and TVD using an appropriate
top hat distribution, for various types of noise and noise rate $p$.

\begin{figure}
    \begin{center}
	\resizebox{0.6\columnwidth}{!}{\includegraphics{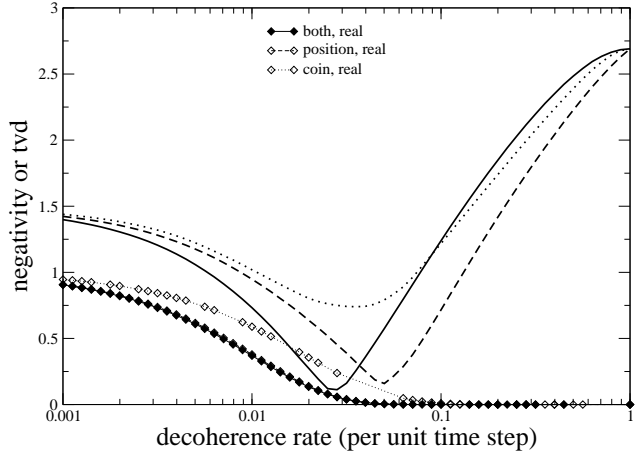}}
    \end{center}
    \caption{Negativity (diamonds) and TVD for
	noise applied to position (dashed), coin (dotted) and both
	(solid) for a quantum walk on the line of 100 steps.}
    \label{fig:ld100bw}
\end{figure}

Guided by the results in \cite{kendon02c} and \cite{carneiro05a},
we focused on the entanglement at the end of the quantum walk,
just before the final measurement,
and considered how it varies as the noise rate $p$ is varied.  We examined
three cases of noise, applying it to both the coin and position, and also
separately to just the coin or the position.
An example is shown in fig.~\ref{fig:ld100bw}.
Both the TVD from the optimal top hat distribution,
and the negativity are plotted.  The minimum in the TVD
indicates the optimal measurement rate $p$.  For a quantum walk of 100 steps
the optimal $p$ ranges through $0.025\lesssim p\lesssim 0.05$.
Note that the minimum for measurements on the coin only is much shallower.
The distribution in this case retains a cusp shape \cite{kendon02c}.
The negativity also remains above zero for longer than when noise is applied
to the position of the walker, indicating the coin measurements are less
effective at removing the quantum correlations.
If noise is applied to the position, with or without noise on the
coin as well, the negativity drops to zero at $p\simeq0.055$,
shortly after the optimal noise rate is reached.
The optimal amount of randomness is thus just about the amount required to
remove all the the quantum correlations from the system.
This make intuitive sense:
we are trying to achieve a uniform distribution on the line, in which any
location within the top hat region is equally likely.  Quantum correlations
distort this smooth distribution, giving it peaks and troughs, especially
at the ends of the top hat \cite{nayak00a,ambainis01a}.
If the noise rate is turned up until the classical random walk is
obtained for $p=1$, classical correlations build up to produce the
binomial distribution in which the quantum walker is more likely to be
found nearer the starting point of the walk.

\section{Entanglement in a quantum walk on cycles}
\label{sec:cyclent}

For the mixing time on cycles, there are extra considerations.  The mixing
time in general depends not only on the size of the cycle, but also on how
close to uniform one sets the threshold $\epsilon$.
As already noted, pure quantum walks don't mix unless something is done to
disrupt the pure quantum evolution, and random or even regular repeated
measurements efficiently change the behaviour into that of fast mixing
to the uniform distribution.  Note also that the optimal rate of measurement
$2\lesssim pN\lesssim 5$ is independent of the threshold $\epsilon$,
even though the main effect is to provide logarithmic scaling of the
mixing time with $\epsilon$.

\begin{figure}
    \begin{center}
        \resizebox{0.8\columnwidth}{!}{\rotatebox{-90}{\includegraphics{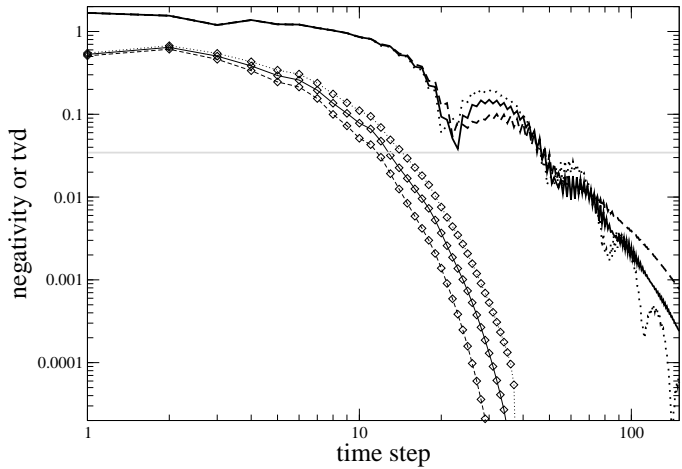}}}
    \end{center}
    \caption{Negativity (diamonds) and TVD for
        noise applied to position for a quantum walk on the
	cycle of size 29 with $p=0.2239$ (dotted),
	$p=0.2511$ (solid) and $p=0.2818$ (dashed).
	The grey horizontal line is at $1/29$.}
    \label{fig:cd29bw}
\end{figure}

We studied how the entanglement varies during the quantum walk
on a cycle, with the noise rate $p$ chosen to be near-optimal.
As with the walk on the line, we examined the three cases
of noise applied to both the position and coin, and applied to just
the position or coin separately.  Again noise applied to the coin
only does not provide a significant improvement in the behaviour compared
with noise applied to the position. 
The results for a typical example, a cycle of size $N=29$ with noise
applied to the position only, are shown in fig.~\ref{fig:cd29bw}.
While the actual mixing time is determined by the choice of $\epsilon$, 
we have indicated the position of $1/N$ in fig.~\ref{fig:cd29bw},
this being the probability of finding the walker at one location
in a uniform distribution.
The time at which the TVD drops below this line is around the time the
entanglement also drops to zero.  The variability of the TVD shows that the
mixing time is not a smooth function of $\epsilon$, so we cannot expect
to determine a more precise result.  As we argued for the walk on the line,
quantum correlations are incompatible with the result we want, a uniform
distribution, so the optimal quantum walk arranges for the quantum
correlations to be removed by the end of the process.

\section{Discussion}
\label{sec:conc}

The optimal quantum walk on the line and cycle is
a carefully balanced combination of quantum dynamics with randomness 
provided by repeated measurements during the evolution of the walk.
We have shown that, while the quantum correlations are necessary
to obtain linear spreading and mixing times, they must be neutralised
to produce a uniform final distribution.
We have also analysed the resources required for a non-unitary quantum
walk and shown that the additional randomness required to obtain the
optimal behaviour is only logarithmic in the size of the problem.
Since a quantum computer has randomness as a built in function, this
is a very efficient use of resources compared to performing a classical
random walk, which requires an amount of randomness quadratic in the
size of the problem.

While the relationship between quantum entanglement and shared
randomness is well-studied in quantum communications theory, 
for example, see \cite{bennett01a}, the
relationship between the two types of resources in quantum computation
has received little attention.  The idea of combining measurements with
quantum gates is well-established \cite{jozsa05a},
but little is known about the importance of each contribution to the
power of the computational process.
Note in particular that in the cluster state model, random measurement
outcomes are corrected, requiring extra computational steps, whereas
the non-unitary quantum walk takes advantage of this randomness to
enhance the computational efficiency.
Our results contribute to the task of quantifying this
relationship and to furthering our understanding of how to harness
the computational power of quantum systems.


We thank many people for interesting discussions of quantum walks,
among them,
Ivens Carneiro\footnote{
We learned with much sadness of his untimely death in a road accident
in April 2006.
},
Hilary Carteret,
Jochen Endrejat,
Barbara Kraus,
Peter Richter,
Barry Sanders,
Mario Szegedy,
and
Tino Tamon
stimulated our thinking for the work in this paper.
VK is funded by a Royal Society University Research Fellowship.



\renewcommand{\baselinestretch}{1.0}\small\normalsize
\bibliography{../bibs/qrw,../bibs/qit,../bibs/shor,../bibs/ent}



\end{document}